\documentclass[prl,aps,twocolumn,showpacs]{revtex4}
\usepackage{graphics}

\begin{document}
\title{Instantaneous Measurement of Non-local Variables}

\author{ Lev Vaidman}


\affiliation 
{ School of Physics and Astronomy\\
Raymond and Beverly Sackler Faculty of Exact Sciences\\
Tel-Aviv University, Tel-Aviv 69978, Israel}

\date{}

\vspace{.4cm}
 
\begin{abstract}
  It is shown, under the assumption of possibility to perform an
  arbitrary local operation, that all nonlocal variables related to
  two or more separate sites can be measured instantaneously, except
  for a finite time required for bringing to one location the
  classical records from these sites which yield the result of the
  measurement.
 It is
  a verification measurement: it yields reliably the eigenvalues of
  the nonlocal variables, but it does not prepare the eigenstates of
  the system.
\end{abstract}


\maketitle

  Seventy years ago Landau and Peierls \cite{LP} claimed that the
  instantaneous measurability of nonlocal variables (i.e., variables
  which related to more than one small  region of space)
  contradicts relativistic causality. Twenty years ago, Aharonov and
  Albert \cite{AA} showed that some nonlocal variables (e.g., the Bell
  operator, see below) can be measured instantaneously and that this
  does not contradict causality. They also showed explicitly how the
  possibility of performing instantaneous von Neumann measurements of
  some other nonlocal variables does contradict causality.  The
  question: ``What are the {\it observables} of relativistic quantum
  theory?'' remains topical even today \cite{BGNP}.
  
  A variable can obtain the status of an observable if it can be
  measured. However, the standard (von Neumann) definition of quantum
  measurement is too restrictive for defining a physical observable:
  the von Neumann definition requires that eigenstates of the measured
  variable are not changed  due to the measurement process.
  The existence of a {\em verification} measurement which yields the
  eigenvalue of a variable with certainty, if prior to the measurement
  the quantum system was in the corresponding eigenstate, is enough
  for giving the status of an observable for such a variable, even if
  the measurement does not leave the system in this eigenstate as the
  von Neumann measurement does.  (If, initially, the system is in a
  superposition or mixture of the eigenstates of the observable, then
  the verification measurement yields one of the corresponding
  eigenvalues according to the quantum probability law.)

  The meaning of ``instantaneous measurement'' is that
in a particular Lorentz frame, at time $t$, observers perform local actions
for a duration of time which can be as short as we wish.
At the end of the measurement
interactions,
 the information about the outcome of the measurement is
classically recorded in the results of local (irreversible) measurements.
In order to  infer which eigenvalue of the nonlocal
variable the system had  originally,  or to generate correctly
distributed probabilistic outcome, these classical results
are later combined at a point within the future light cones of all the
observers. 

Note the difference with the case of {\it exchange measurements}
\cite{AAV} which can also be performed for all nonlocal variables. In
the exchange measurement, local  operations of swapping lead to
swapping between the quantum state of the composite system and the
quantum state of the local separated parts of the measuring device. In
order to find out which eigenvalue the system had originally, it is
required coherent maintaining of all these parts until they enter the
forward light cone of all of the original subsystems one wishes to
measure where final  local joint measurement is performed.  After instantaneous swapping, the outcome of the measurement is not written in the form of
classical information and, in fact, the outcome of the quantum
measurement does not exist yet: at this stage the exchange measurement
can be reversed and the system can be brought back to its original (in
general unknown) state.

In this Letter, I will show that apart from variables related to the spread-out
fermionic  wave function, {\it all} nonlocal variables have the status of
observables in the framework of relativistic quantum mechanics, i.e.,
all variables related to two or more separate sites are measurable
instantaneously using verification measurements.  This includes
variables with entangled eigenstates and nonlocal variables with
product eigenstates \cite{benett}.

Verification measurements have been considered before. It has been
shown \cite{PV} that verification measurements of some nonlocal
variables erase local information and, therefore, cannot be ideal von
Neumann measurements. Recently, Groisman and Reznik \cite{GR} showed
that there are instantaneous verification measurements for all spin
variables of a system of two separated spin-$1\over 2$ particles. Consider, for example a nonlocal variable of two  spin-$1\over 2$ particles located in
separate locations $A$ and $B$, whose
eigenstates are the following
product states:
\begin{eqnarray}
\label{4g}
|\Psi_1\rangle=&|\uparrow_z\rangle_A&|\uparrow_z\rangle_B,\nonumber\\
|\Psi_2\rangle=&|\uparrow_z\rangle_A&|\downarrow_z\rangle_B,\\
|\Psi_3\rangle=&|\downarrow_z\rangle_A&|\uparrow_x\rangle_B,\nonumber\\
|\Psi_4\rangle=&|\downarrow_z\rangle_A&|\downarrow_x\rangle_B.\nonumber
\end{eqnarray}

An instantaneous  ideal von Neumann measurement of
this variable does contradict causality. Assume that at time $t$ such
an ideal measurement is performed. Then we can send superluminal signal
from $A$ to $B$ in the following way. We prepare in advance the system
in the state $|\Psi_1\rangle$ and agree that Bob at site $B$ measures
the spin $z$ component of his particle shortly after time $t$. Now, in order to send a
superluminal signal, Alice at site $A$ can at a very short time before
time $t$  flip her spin. If she does so, then after the nonlocal
measurement at time $t$, the system will end up either in state
$|\Psi_3\rangle$ on in state $|\Psi_4\rangle$. In both cases Bob has a
nonvanishing probability to find his spin ``down'' in $\hat z$ direction, while this
probability is zero if Alice decides not to flip her spin.

The method for  the verification measurement
I present here
uses teleportation technique
\cite{BBGCJPW}.
The first step  is the teleportation of the state of
the spin from $B$ (Bob's site) to $A$ (Alice's site). Bob and Alice do
not perform the full teleportation (which invariably requires a finite
period of time), but only the Bell measurement at Bob's site which might
last, in principle, as short a time as we wish. (I will continue to use
the term ``teleportation'' just for this first step of the original
proposal \cite{BBGCJPW}.)
 
In the teleportation procedure for a spin-$1\over 2$ particle we start with
a prearranged EPR(Bohm)  pair   of spin-$1\over 2$ particles
one of which is located at Bob's site and another at Alice's site, $|\Psi_-\rangle_{AB}={1\over \sqrt 2}(|{\uparrow}\rangle_A
|{\downarrow}\rangle_B - |{\downarrow}\rangle_A
|{\uparrow}\rangle_B)$.  The
procedure is based on the identity
\begin{eqnarray}
  \label{ident}
  |\Psi\rangle_1|\Psi_-\rangle_{2,3} = {1\over 2}
  (|\Psi_-\rangle_{1,2} | \Psi\rangle_{3} + 
|\Psi_+\rangle_{1,2} |\tilde \Psi^{(z)}\rangle_{3} +\nonumber \\
|\Phi_-\rangle_{1,2} |\tilde \Psi^{(x)}\rangle_{3}+
|\Phi_+\rangle_{1,2} |\tilde \Psi^{(y)}\rangle_{3}),
\end{eqnarray}
where $|\Psi_\mp\rangle = {1\over \sqrt 2}(|{\uparrow}\rangle
|{\downarrow}\rangle \mp |{\downarrow}\rangle |{\uparrow}\rangle)$, $|\Phi_\mp\rangle = {1\over \sqrt 2}(|{\uparrow}\rangle
|{\uparrow}\rangle \mp |{\downarrow}\rangle |{\downarrow}\rangle)$ are
eigenstates of the Bell operator and  $|\tilde \Psi^{(z)}\rangle$
signifies the state  $| \Psi\rangle$ 
 rotated  by $\pi$ around  $\hat z$ axis, etc.
Thus, the Bell operator measurement performed on the two particles in
 Bob's site ``collapses'' (or effectively
collapses) to one of the branches of the superposition, the RHS of (\ref{ident}), and, therefore, 
 teleports the state $ |\Psi\rangle$ of Bob's particle to Alice except for a possible rotation by $\pi$ (known to Bob) around one
of the axes.

The second step is taken by Alice. She can perform it at time $t$
without waiting for Bob. She measures the spin of her particle in the
$z$ direction. If the result is ``up'', she measures the spin of the
particle teleported from Bob in the $z$ direction, and if her spin is
``down'', she measures the spin of the Bob's particle in the $x$
direction.

 This completes the  measurement except for combining local results
 together for finding out the result of the nonlocal measurement.
Indeed, the eigenstates of the spin in the $z$ direction and in the
$x$ direction  are
teleported without leaving their  lines. Thus, Bob's knowledge about
possible flip together with Alice's results distinguish unambiguously
between the four eigenstates (\ref{4g}).

The method I presented above can be modified for measurement of other
nonlocal variables of two spin-$1\over 2$ particles.  However, I will
turn now to another, universal, method which is applicable to any
nonlocal variable $O(q_A, q_B, ...)$, where $q_A$ belongs to region
$A$, etc. I will not try to optimize the method or consider any
realistic proposal: my task is to show that, given unlimited resources
of entanglement and arbitrary local interactions, any nonlocal
variable is measurable.

I will start with the case of a general variable of a composite
systems with two parts. First, (for simplicity), Alice and Bob perform
unitary operations which swap  the states of their systems with the  states of
sets of K spin-$1\over 2$ particles. In this way Alice and Bob will need
the teleportation procedure for spin-$1\over 2$ particles
only. Teleportation of the states  of all $K$ individual  spins leads to teleportation the state of the
 set, be it  entangled or not.

 The general protocol is illustrated in Fig. 1. The resources include
 numerous teleportation channels arranged in a particular way: two
 channels for the first round of back and forth teleportations, then
 $4^K-1$ clusters,  each includes two channels for the second round of back and
 forth teleportations and $4^{2K}-1$ sub-clusters. Each sub-cluster, in turn, includes
 two channels for the third round of teleportation and  $4^{2K}-1$ similar
 sub-sub-clusters, etc.  The protocol consists of the following steps:

\begin{figure*}
\includegraphics{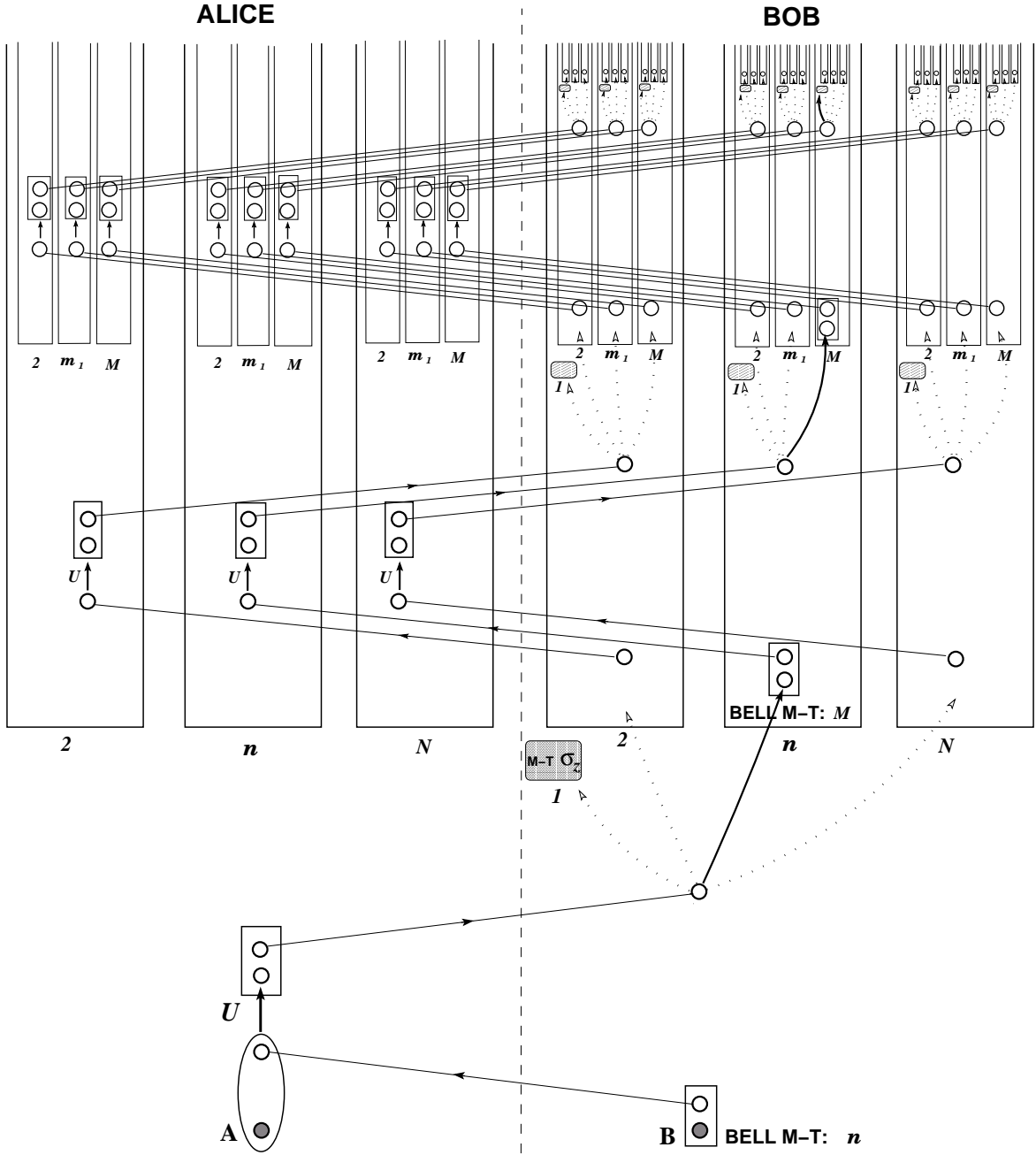}
\caption{The scheme of the measurement of a nonlocal
  variable
 of a two-part system. In the example shown on the figure, 
In the  example shown on the figure, the results of
  the Bell measurement in Bob's site were $n, M, 1$. Thus, the nonlocal
  measurement has been essentially completed after three teleportation rounds.}
\end{figure*}

$\bullet$    Bob teleports his system ($K$ spin-$1\over 2$ particles) to Alice and records the outcome of
the Bell measurements $n$. 

As before, ``teleports'' means that Bob performs the Bell
measurements but does not send the outcome to Alice.  The number of
possible outcomes is $N = 4^K$. We signify them by $n=1,2,...N$, with
$n=1$ corresponding to singlets in all Bell measurements, i.e. to  teleportation  without distortion.

$\bullet$  Alice performs a unitary operation $U$ on the composite system of her
 and the teleported spins which, under the assumption of
non-distorted teleportation, transforms the eigenstates of the nonlocal
variable (which now actually are fully located in Alice's site) to
product states in which each spin is either ``up'' or ``down'' along
the $z$ direction.

$\bullet$  Alice  teleports the complete composite
system ($2K$ spin-$1\over 2$ particles) to Bob.

 Note that 
if the system is  in one of the product states
in the spin $z$ basis, then it will remain in this basis.

$\bullet$  If $n=1$ Bob measures the teleported system in the spin $z$
basis.

In this case  (the probability for which is $1\over N$), Bob gets the
composite system in one of the spin $z$ product states and his
measurements in the spin $z$ basis complete the
measurement of the nonlocal variable.

If  $n\neq 1$ Bob teleports the system back to Alice in the
teleportation channel of cluster $n$. He records the outcome of
the Bell measurements $m_1$ which can have values from $1$ to $M=4^{2K}$.

Since in this case Alice's operations
do not bring the eigenstates of the nonlocal variable to the spin
$z$ basis,  Bob teleports the
system back to Alice ``telling'' her the outcome of his previous Bell
measurements via the channel he uses for the teleportation.

$\bullet$ Alice performs unitary operations on each system in $N-1$
teleportation channels of the second round which, under the assumption
of no distortion in these teleportations, transforms the
eigenstates of the nonlocal variable to product spin $z$ eigenstates.

Alice's operations include corrections required due to her and Bob's teleportations  and her unitary transformation of the first round.

$\bullet$   Alice  teleports all $N-1$ 
 systems back to Bob.

$\bullet$  If $m_1=1$ Bob measures the system teleported from Alice in
cluster $n$  in the spin $z$
basis.

Again, in that case, the spin measurements complete the measurement
of the nonlocal variable, since their results together with the
outcomes of Alice's and Bob's Bell measurements specify uniquely the
eigenvalue of the nonlocal variable.

If  $m_1\neq 1$ Bob teleports the system back to Alice in the
teleportation channel of sub-cluster $m_1$ of cluster $n$. He records the outcome of
the Bell measurements $m_2$.

$\bullet$  Alice performs  unitary operations on each system in $(N-1)(M-1)$
teleportation channels of the third round. The operation on each
system is such that
if Bob, indeed, teleported the system in this channel, and  if his last
teleportation happened to be without distortion, then the eigenstates
of the nonlocal variable are transformed into
product spin $z$ states.

Alice's operations include corrections required due to her and Bob's teleportations  and her unitary transformations of the first and second rounds.

 $\bullet$  Alice  teleports all $(N-1)(M-1)$ 
systems back to Bob.

 $\bullet$ If $m_2=1$ Bob measures the system teleported from Alice in
 sub-cluster $m_1$ of cluster $n$ in the spin $z$ basis.

If  $m_2\neq 1$ Bob teleports the system back to Alice in the
teleportation channel of sub-sub-cluster $m_2$ of sub-cluster $m_1$ of cluster $n$. He records the outcome of
the Bell measurements $m_3$.

Alice and Bob continue this procedure.  The nonlocal measurement is
completed when, for the first time, Bob performs a teleportation
without distortion.  Since, conceptually, there is no limitation for
the number of teleportation rounds, and each round (starting form the
second) has the same probability for success, $1\over M$, the
measurement of the nonlocal variable can be performed with probability
arbitrarily close to 1. Given the desired probability of the
successful nonlocal measurement, Alice and Bob decide about the number
of rounds of teleportations.  The number of entangled pairs required
for each round grows exponentially with the number of rounds. While
Bob uses only one teleportation channel in each round and stops after
his first teleportation without distortion, Alice has to perform all
teleportations in all channels.

The generalization to a system with more than two parts is more or
less straightforward. Let us sketch it for three-part system. First,
Bob and Carol teleport their parts to Alice. Alice performs a unitary
transformation which, under the assumption of undisturbed
teleportations of both Bob and Carol, transforms the eigenstates of the
nonlocal variable to product states in the spin $z$ basis. Then she
teleports the complete system to Bob. Bob teleports it to Carol in a
particular channel $n_B$ depending on the results of the Bell
measurement of his first teleportation. Carol teleports all the systems
from the teleportation channels from Bob back to Alice. In particular,
the system from channel $i_B$ she teleports in the channel $(n_B, n_C)$ depending
on her Bell measurement result $n_C$. The system corresponding to $(n_B,
n_C)= (1,1)$ is not teleported, but measured by Carol in the spin $z$
basis. Alice knows the transformation performed on the system which arrives
in her channels $(n_B, n_C)$  except for corrections due to the last
teleportations of Bob and Carol. She  assumes that there were
no distortion in those, and teleports all the systems back to Bob after the
unitary operation which transforms the eigenstates of the 
variable to product states in the spin $z$ basis.  Alice, Bob and Carol
continue the procedure until the desired probability of successful
measurement is achieved.

The required resources, such as the number of teleportation channels
and required number of operations are very large, but this does not
concern us here. We have shown that there are no relativistic
constraints preventing instantaneous measurement of any variable of a
quantum system with spatially separated parts, answering the above
long standing question. 
This question is relevant for quantum cryptography and quantum
computation performed with distributed systems. The practical
advantage of the method presented in this Letter is that it relies on prior entanglement
 and does not require coherent
transportation of quantum systems.

Can this result be generalized to a quantum system which  itself is
in a superposition of being in different places? The key to this
question is the generality of the assumption of the possibility to
perform any local operation. If a quantum state of a particle which is
in a nonlocal superposition  can be locally transformed to
(an entangled) state of local quantum systems, then any variable of the
particle is measurable through the  measurement of the
corresponding composite system. However, while for bosons it is clear
that there are such local operations (transformation of photon state
to entangled state of atoms has been achieved in the laboratory
\cite{Har}), for fermion states the situation is different \cite{AV}.
If the transformation of a superposition of a fermion state to local
variables is possible, then these local separated in space variables
should fulfill anti-commutation relations. This is the reason to expect
super-selection rules which prevent such transformations.

I am grateful for  Yakir Aharonov, Charles Bennett, Shmuel Nussinov, and Benni
Reznik  for useful
comments. This research was supported in part by grant 62/01 of the
Israel Science Foundation and by the  MOD 
 Unit.



\end{document}